\documentstyle[graphicx]{mn2e}

\begin{document}

\title[WiggleZ Survey: Alcock-Paczynski measurement]{The WiggleZ Dark
  Energy Survey: measuring the cosmic expansion history using the
  Alcock-Paczynski test and distant supernovae}

\author[Blake et al.]{\parbox[t]{\textwidth}{Chris
    Blake$^1$\footnotemark, Karl Glazebrook$^1$, Tamara
    M.\ Davis$^{2,3}$, Sarah Brough$^4$, \\ Matthew Colless$^4$,
    Carlos Contreras$^1$, Warrick Couch$^1$, Scott Croom$^5$,
    \\ Michael J.\ Drinkwater$^2$, Karl Forster$^6$, David
    Gilbank$^7$, Mike Gladders$^8$, \\ Ben Jelliffe$^5$, Russell
    J.\ Jurek$^9$, I-hui Li$^1$, Barry Madore$^{10}$,
    \\ D.\ Christopher Martin$^6$, Kevin Pimbblet$^{11}$, Gregory
    B.\ Poole$^1$, Michael Pracy$^{1,12}$, \\ Rob Sharp$^{2,12}$,
    Emily Wisnioski$^1$, David Woods$^{13}$, Ted K.\ Wyder$^6$ and
    H.K.C. Yee$^{14}$} \\ \\ $^1$ Centre for Astrophysics \&
  Supercomputing, Swinburne University of Technology, P.O. Box 218,
  Hawthorn, VIC 3122, Australia \\ $^2$ School of Mathematics and
  Physics, University of Queensland, Brisbane, QLD 4072, Australia
  \\ $^3$ Dark Cosmology Centre, Niels Bohr Institute, University of
  Copenhagen, Juliane Maries Vej 30, DK-2100 Copenhagen \O, Denmark
  \\ $^4$ Australian Astronomical Observatory, P.O. Box 296, Epping,
  NSW 1710, Australia \\ $^5$ Sydney Institute for Astronomy, School
  of Physics, University of Sydney, NSW 2006, Australia \\ $^6$
  California Institute of Technology, MC 278-17, 1200 East California
  Boulevard, Pasadena, CA 91125, United States \\ $^7$ Astrophysics
  and Gravitation Group, Department of Physics and Astronomy,
  University of Waterloo, Waterloo, ON N2L 3G1, Canada \\ $^8$
  Department of Astronomy and Astrophysics, University of Chicago,
  5640 South Ellis Avenue, Chicago, IL 60637, United States \\ $^9$
  Australia Telescope National Facility, CSIRO, Epping, NSW 1710,
  Australia \\ $^{10}$ Observatories of the Carnegie Institute of
  Washington, 813 Santa Barbara St., Pasadena, CA 91101, United States
  \\ $^{11}$ School of Physics, Monash University, Clayton, VIC 3800,
  Australia \\ $^{12}$ Research School of Astronomy \& Astrophysics,
  Australian National University, Weston Creek, ACT 2611, Australia
  \\ $^{13}$ Department of Physics \& Astronomy, University of British
  Columbia, 6224 Agricultural Road, Vancouver, BC V6T 1Z1, Canada
  \\ $^{14}$ Department of Astronomy and Astrophysics, University of
  Toronto, 50 St.\ George Street, Toronto, ON M5S 3H4, Canada}

\maketitle

\begin{abstract}
Astronomical observations suggest that today's Universe is dominated
by a dark energy of unknown physical origin.  One of the most notable
consequences in many models is that dark energy should cause the
expansion of the Universe to accelerate: but the expansion rate as a
function of time has proven very difficult to measure directly.  We
present a new determination of the cosmic expansion history by
combining distant supernovae observations with a geometrical analysis
of large-scale galaxy clustering within the WiggleZ Dark Energy
Survey, using the Alcock-Paczynski test to measure the distortion of
standard spheres.  Our result constitutes a robust and non-parametric
measurement of the Hubble expansion rate as a function of time, which
we measure with $10$-$15\%$ precision in four bins within the redshift
range $0.1 < z < 0.9$.  We demonstrate that the cosmic expansion is
accelerating, in a manner independent of the parameterization of the
cosmological model (although assuming cosmic homogeneity in our data
analysis).  Furthermore, we find that this expansion history is
consistent with a cosmological-constant dark energy.
\end{abstract}

\section{Introduction}
\renewcommand{\thefootnote}{\fnsymbol{footnote}}
\setcounter{footnote}{1}
\footnotetext{E-mail: cblake@astro.swin.edu.au}

Observations by astronomers over the past fifteen years have suggested
that the Universe is dominated by an unexpected component known as
``dark energy'', yet we still have no physical understanding of its
existence or magnitude.  Determining the nature of dark energy is one
of the most important challenges for contemporary cosmology because
its presence implies that our understanding of the physics of the
Universe is incomplete.  Possible explanations are that the theory of
General Relativity must be modified on cosmological scales to generate
an effective repulsive gravitational force, the Universe is filled
with a diffuse material that acts with an effective negative pressure,
or our interpretation of cosmological observations must be changed to
reflect inhomogeneities.

One of the most important consequences of the first two scenarios is
that the Universe underwent a transition from decelerating to
accelerating expansion within the last few billion years.  In this
paper we use new observations and methods to map this expansion
history in a model-independent, non-parametric and robust manner.  We
do assume cosmic homogeneity in our data analysis, and therefore we do
not address the third scenario.

In the standard cosmological analyses, involving datasets such as
distant supernovae (Riess et al.\ 1998, Perlmutter et al.\ 1999,
Amanullah et al.\ 2010, Conley et al.\ 2011), galaxy surveys
(Giannantonio et al.\ 2008, Percival et al.\ 2010, Blake et
al.\ 2011b) and the Cosmic Microwave Background radiation (Komatsu et
al.\ 2011), the expansion rate of the Universe at different look-back
times is not measured directly.  Accelerating expansion is a {\it
  model-dependent implication of fitting cosmological data with
  parametric models} in which prior assumptions are made about the
physical components of the Universe and their evolution with redshift.
However, the unknown physical nature of dark energy implies that we
cannot yet be certain that any particular parametric model is correct.

For example, measurements of distant supernovae provide some of the
strongest evidence for the existence of dark energy.  The observed
quantities are the apparent magnitudes of the supernovae, which yield
a relative luminosity distance as a function of redshift,
i.e.\ $D_L(z) \, H_0/c$ where $D_L(z)$ is the luminosity distance,
$H_0$ is the local value of the Hubble expansion rate and $c$ is the
speed of light.  These supernovae distances are well-fit by a model in
which the physical contents of the Universe are divided into
pressureless matter with a current fractional energy density
$\Omega_{\rm m}$, which dilutes with redshift as $(1+z)^3$, and a
``cosmological constant'' $\Omega_\Lambda$ whose energy density does
not change with redshift.  The best-fitting parameters assuming this
model are $\Omega_{\rm m} \approx 0.27$ and $\Omega_\Lambda \approx
0.73$, implying a transition from decelerating to accelerating
expansion at $z \approx 0.6$ (e.g.\ Conley et al.\ 2011).

However, this conclusion is dependent on the assumed parameterized
model.  The acceleration of the expansion rate is not directly
observed.  When the supernovae data are subject to model-independent
{\it non-parametric analysis}, the evidence for accelerating expansion
is much weaker (Wang \& Tegmark 2005, Shapiro \& Turner 2006,
Shafieloo 2007, Sollerman et al.\ 2009, Shafieloo \& Clarkson 2010).
This is primarily because the acceleration rate is obtained as the
second derivative of the noisy observed luminosity distances.
Moreover, obtaining the cosmic expansion rate from a luminosity
distance requires an additional assumption about spatial curvature.
Although recent cosmological data is consistent with a spatially-flat
Universe (Percival et al.\ 2010, Komatsu et al.\ 2011), this
conclusion is again reached within the assumption of a parameterized
model (unless we invoke a strong prior from inflationary cosmological
models).

The expansion history of the Universe as a function of redshift $z$ is
described by the evolution of the Hubble parameter $H(z) \equiv (1+z)
\, da/dt$, where $a(t)$ is the cosmic scale factor at time $t$.  In
this paper we present a new determination of the function $H(z)$ using
a two-step approach.  Firstly we apply an Alcock-Paczynski measurement
(Alcock \& Paczynski 1979) to the large-scale clustering of galaxies
in the WiggleZ Dark Energy Survey (Drinkwater et al.\ 2010), thereby
measuring the distortion parameter $F(z) \equiv (1+z) D_A(z) H(z)/c$,
where $D_A(z)$ is the physical angular-diameter distance.  Secondly we
combine this measurement with the supernovae luminosity distances
$D_L(z) \, H_0/c$, using the equivalence of distance measurements
$D_L(z) = D_A(z) \, (1+z)^2$, to infer the value of $H(z)/H_0$.  The
result is an accurate non-parametric reconstruction of the cosmic
expansion history, which is independent of the assumed cosmological
model (including spatial curvature).  We present our results in the
form of both individual measurements of $H(z)/H_0$ in four redshift
bins within the redshift range $0.1 < z < 0.9$, and a non-parametric
reconstruction of a continuous cosmic expansion history using a
modified version of the iterative methods introduced by Shafieloo et
al.\ (2006).

Other methods for determining the Hubble expansion rate have been
previously attempted.  Firstly, $H(z)$ may be inferred from the
relative ages of passively-evolving galaxies (Stern et al.\ 2010).
This is a promising technique although subject to many assumptions
about the stellar populations and formation epochs of these galaxies.
Secondly, the imprint of baryon acoustic oscillations in galaxy
clustering has been applied as a standard ruler along the
line-of-sight (Gaztanaga, Cabre \& Hui 2009).  However, the current
availability of large-scale galaxy survey data restricts this latter
technique to low redshift $z=0.35$ and it is debatable whether the
signal-to-noise of current measurements is in fact sufficient to
ensure a robust measurement (Kazin et al.\ 2009).  Finally, the
physical presence of dark energy has been inferred from the late-time
Integrated Sachs-Wolfe effect which correlates low-redshift galaxy
overdensities with Cosmic Microwave Background (CMB) anisotropies
(Giannantonio et al.\ 2008).  However, this last technique does not
allow recovery of the detailed expansion history and the achievable
statistical significance is limited.

The Alcock-Paczynski test (Alcock \& Paczynski 1979) is a geometric
probe of the cosmological model based on the comparison of the
observed tangential and radial dimensions of objects which are assumed
to be isotropic in the correct choice of model.  It can be applied to
the 2-point statistics of galaxy clustering if redshift-space
distortions, the principal additional source of anisotropy, can be
successfully modelled (Ballinger, Peacock \& Heavens 1996, Matsubara
\& Suto 1996, Matsubara 2000, Seo \& Eisenstein 2003, Simpson \&
Peacock 2010).  This method has been previously utilized with data
from the 2-degree Field Quasar Survey (Outram et al.\ 2004), but the
WiggleZ survey offers a superior measurement owing to the reduced
importance of shot noise.  Chuang \& Wang (2011) recently performed
fits to the tangential and radial clustering of Luminous Red Galaxies
within the Sloan Digital Sky Survey (SDSS) at redshift $z=0.35$,
utilizing Alcock-Paczynski information.  Indeed, a general analysis of
the tangential/radial galaxy clustering pattern in the presence of
baryon acoustic oscillations demonstrates how the information may be
divided into an overall scale distortion, quantified by a distance
parameter $D_V \propto (D_A^2/H)^{1/3}$, and a warping, quantified by
the Alcock-Paczynski distortion factor $D_A H$, enabling the
disentangling of $D_A$ and $H$ (Padmanabhan \& White 2008; Taruya,
Saito \& Nishimichi 2011; Kazin, Sanchez \& Blanton 2011).  Such
approaches are just becoming possible with the current generation of
large-scale galaxy surveys, and will be very powerful when applied to
future datasets such as the Baryon Oscillation Spectroscopic Survey
(BOSS, Eisenstein et al.\ 2011).  Another recent application of the
Alcock-Paczynski test was presented by Marinoni \& Buzzi (2010), who
presented a study of the distribution of close galaxy pairs.

The WiggleZ Dark Energy Survey (Drinkwater et al.\ 2010) at the
Australian Astronomical Observatory has recently provided a new
large-scale galaxy redshift survey dataset, allowing low-redshift
cosmological measurements in the SDSS to be extended to higher
redshifts up to $z=0.9$.  In particular, the survey has yielded a
measurement of the baryon acoustic peak in the clustering pattern at
$z=0.6$ (Blake et al.\ 2011b).  In addition, Blake et al.\ (2011a)
mapped the growth of cosmic structure in this dataset across the
redshift range $0.1 < z < 0.9$ for a fixed background cosmological
model.  The new study presented in this paper constitutes a
generalization of that analysis incorporating Alcock-Paczynski
distortions arising in varying cosmological models.

Our paper is structured as follows: in Section \ref{secwigglez} we
introduce the WiggleZ survey dataset and power spectrum measurements.
In Section \ref{secap} we describe our implementation of the
Alcock-Paczynski test and our model for marginalizing over
redshift-space distortions.  In Section \ref{secapsn} we combine these
Alcock-Paczynski measurements with supernovae data to deduce the
expansion history $H(z)$.  In Section \ref{secrecon} we present a
non-parametric reconstruction of the continuous cosmic expansion
history, and we summarize our findings in Section \ref{secconc}.

\section{WiggleZ Survey power spectra}
\label{secwigglez}

\subsection{The WiggleZ Dark Energy Survey}

The WiggleZ Dark Energy Survey (Drinkwater et al.\ 2010) at the
$3.9$-m Anglo-Australian Telescope has provided the next step in
large-scale spectroscopic galaxy redshift surveys, mapping a cosmic
volume $\sim 1$ Gpc$^3$ over the redshift range $0 < z < 1$.  By
covering a total of about $800$ deg$^2$ of sky the WiggleZ survey has
mapped about 100 times more effective cosmic volume in the $z > 0.5$
Universe than previous galaxy redshift surveys.  Target galaxies were
chosen by a joint selection in UV and optical wavebands, using
observations by the Galaxy Evolution Explorer satellite (GALEX; Martin
et al.\ 2005) matched with ground-based optical imaging from the Sloan
Digital Sky Survey (York et al.\ 2000) in the Northern Galactic Cap,
and from the Red-sequence Cluster Survey 2 (RCS2) (Gilbank et
al.\ 2011) in the Southern Galactic Cap.  A series of magnitude and
colour cuts (Drinkwater et al.\ 2010) was used to preferentially
select high-redshift star-forming galaxies with bright emission lines,
that were then observed using the AAOmega multi-object spectrograph
(Sharp et al.\ 2006) in 1-hr exposures.  In this study we analyzed a
galaxy sample drawn from our final set of observations, after cuts to
maximize the contiguity of each survey region.  The sample includes a
total of $162{,}323$ galaxy redshifts, which we divided into four
different redshift slices of width $\Delta z = 0.2$ between $z=0.1$
and $z=0.9$.  The numbers of galaxies analyzed in each redshift slice
were $N = (22072, 42160, 63737, 34354)$.

\subsection{Power spectrum measurements}

Our analysis is based on measurements of the power spectrum amplitude
of WiggleZ galaxies as a function of the angle to the line-of-sight
$\theta$, which we parameterize by the variable $\mu = \cos{\theta}$.
By comparing the amplitude of power spectrum modes as a function of
$\mu$ we can apply an Alcock-Paczynski measurement as proposed by
Ballinger et al.\ (1996) and Matsubara \& Suto (1996) and described
below in Section \ref{secap}.  We measured galaxy power spectra in six
independent survey regions, dividing each region into four redshift
slices.  At a given redshift we fitted models to the measurements in
each different survey region, convolving with the respective survey
window function.

In order to measure the power spectrum and calculate this convolution
we need to define the selection function of the survey which describes
the mean galaxy density as a function of position.  The angular and
redshift dependences of the selection function for each WiggleZ region
were determined using the methods described by Blake et al.\ (2010).
This process models several effects including the variation of the
GALEX target density with dust and exposure time, the incompleteness
in the current redshift catalogue, the variation of that
incompleteness imprinted across each 2-degree field by constraints on
the positioning of fibres and throughput variations with fibre
position, and the dependence of the galaxy redshift distribution on
optical magnitude.  The galaxy power spectrum was then measured using
the optimal-weighting scheme proposed by Feldman, Kaiser \& Peacock
(1994).  Our fiducial cosmological model which we use to convert
redshifts and angular co-ordinates to distances is a flat $\Lambda$CDM
model with matter density $\Omega_{\rm m} = 0.27$.  We therefore
assume cosmic homogeneity in our data analysis.

\begin{figure*}
\begin{center}
\resizebox{16cm}{!}{\rotatebox{270}{\includegraphics{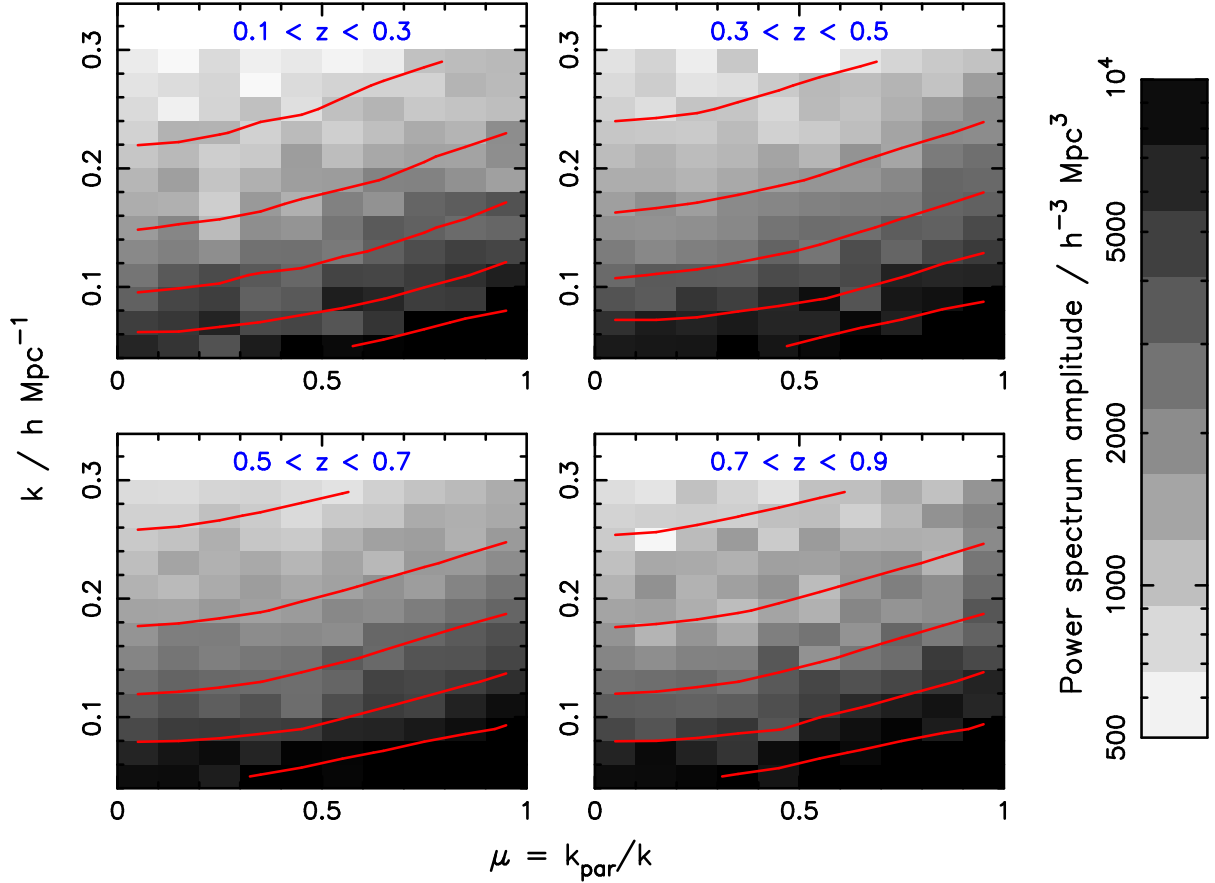}}}
\end{center}
\caption{The galaxy power spectrum amplitude as a function of
  amplitude and angle of Fourier wavevector $(k,\mu)$, determined by
  stacking observations in different WiggleZ survey regions in four
  redshift slices.  The contours correspond to the best-fitting
  non-linear redshift-space distortion model.  We note that because of
  the differing degrees of convolution in each region due to the
  window function, a ``de-convolution'' method was used to produce
  this plot.  Before stacking, the data points were corrected by the
  ratio of the unconvolved and convolved two-dimensional power spectra
  corresponding to the best-fitting model, for the purposes of this
  visualization.  In the absence of redshift-space distortions, the
  model contours would be horizontal lines if the fiducial cosmology
  was equal to the true cosmology.}
\label{figpkmeas}
\end{figure*}

Because the WiggleZ survey consists of a series of independent
narrow-angle cones (of width $\approx 10$ degrees on the sky) a
``flat-sky'' approximation may be used in which the co-ordinate
$x$-axis is aligned with the line-of-sight at the centre of each
region; the wavevector components $(k,\mu)$ are then deduced from the
three-dimensional Fourier vector $\vec{k} = (k_x, k_y, k_z)$ as $k =
\sqrt{k_x^2 + k_y^2 + k_z^2}$ and $\mu = |k_x|/k$.  The observed power
spectra were corrected for the small effect of redshift blunders
(Blake et al.\ 2010).  The covariance matrix of the power spectrum
measurement was deduced using the methodology of Feldman et
al.\ (1994), and the convolution matrices were determined from the
window function as described in Section 2.2 of Blake et al.\ (2011a).
Figure \ref{figpkmeas} displays the 2D power spectrum amplitudes as a
function of $(k,\mu)$, stacked over the survey regions at each
redshift.

The effective redshifts of the power spectrum measurements in each
slice were $z = (0.22, 0.41, 0.6, 0.78)$, determined by weighting each
pixel in the selection function by its contribution to the power
spectrum error (Blake et al.\ 2010):
\begin{equation}
  z_{\rm eff} = \sum_{\vec{x}} z \, \left( \frac{n_g(\vec{x})
      P_g}{1 + n_g(\vec{x}) P_g} \right)^2 ,
\end{equation}
where $n_g(\vec{x})$ is the galaxy number density in each grid cell
$\vec{x}$ and $P_g$ is the characteristic galaxy power spectrum
amplitude, which we evaluated at a scale $k = 0.1 \, h$ Mpc$^{-1}$.
We note that our results are not sensitive to the precise value of the
effective redshift: the measured quantity is the scale distortion
parameter $F(z_{\rm eff}) \equiv (1 + z_{\rm eff}) D_A(z_{\rm eff})
H(z_{\rm eff})$ {\it relative to its value in the fiducial
  cosmological model}, rather than the quantity $F(z_{\rm eff})$
itself.  In other words, if the effective redshift was in error by
$\Delta z_{\rm eff}$, then the systematic error in $F(z_{\rm
  eff}+\Delta z_{\rm eff})/F_{\rm fid}(z_{\rm eff}+\Delta z_{\rm
  eff})$ can be negligible even when the difference between $F(z_{\rm
  eff})$ and $F(z_{\rm eff}+\Delta z_{\rm eff})$ is not.

\section{Alcock-Paczynski measurement}
\label{secap}

\begin{figure*}
\begin{center}
\resizebox{16cm}{!}{\rotatebox{270}{\includegraphics{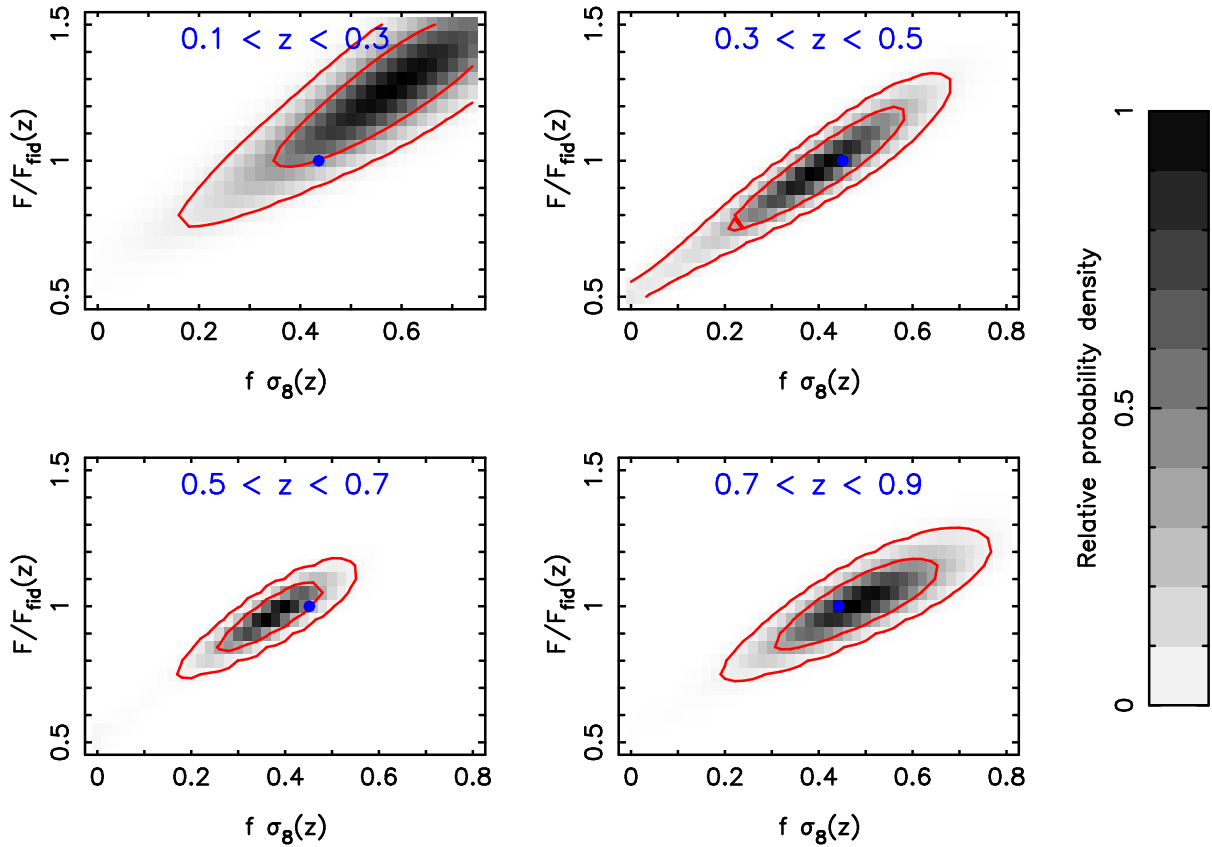}}}
\end{center}
\caption{This Figure displays the joint likelihood of the
  Alcock-Paczynski scale distortion parameter $F(z)$ relative to the
  fiducial value $F_{\rm fid}$, and the growth rate quantified by $f
  \, \sigma_8(z)$, obtained from fits to the 2D galaxy power spectra
  of the WiggleZ Dark Energy Survey in four redshift slices.  In order
  to produce this Figure we marginalized over the linear galaxy bias
  $b^2$ and the pairwise velocity dispersion $\sigma_v$.  There is
  some degeneracy between $F$ and $f \, \sigma_8$ but their
  characteristic dependence on the angle to the line-of-sight is
  sufficiently different that both parameters may be successfully
  extracted.  The probability density is plotted as both greyscale and
  contours enclosing $68\%$ and $95\%$ of the total likelihood.  The
  solid circles indicate the parameter values in our fiducial
  cosmological model.}
\label{figapbeta}
\end{figure*}

\subsection{The Alcock-Paczynski effect}

An Alcock-Paczynski measurement (Alcock \& Paczynski 1979) is a method
for constraining the cosmological model by comparing the observed
tangential and radial dimensions of objects which are assumed to be
isotropic, i.e.\ possess equal co-moving tangential and radial sizes
$L_0$.  The observed tangential dimension is the angular projection
$\Delta \theta = L_0/[(1+z) D_A(z)]$.  The observed radial dimension
is the redshift projection $\Delta z = L_0 H(z)/c$.  The intrinsic
size $L_0$ does not need to be known in order to recover the
observable $\Delta z/\Delta \theta = (1+z) D_A(z) H(z)/c$, which is
independent of any assumption about spatial curvature.  An
Alcock-Paczynski measurement does not necessarily need to be applied
to cosmological ``objects'' but is equally valid for an isotropic
process such as the 2-point statistics of galaxy clustering (Ballinger
et al.\ 1996, Matsubara \& Suto 1996).

The relative tangential/radial distortion depends on the value of
$F(z) = (1+z) D_A(z) H(z)/c$ relative to the fiducial model (which we
take as a flat $\Lambda$ Cold Dark Matter cosmology with $\Omega_{\rm
  m} = 0.27$, although the results do not depend significantly on this
choice).  We applied the Alcock-Paczynski methodology to the WiggleZ
survey clustering by measuring the relative amplitude of power
spectrum modes as a function of their angle to the line-of-sight
(Ballinger et al.\ 1996, Matsubara \& Suto 1996, Simpson \& Peacock
2010), assuming the underlying isotropy of these modes in the true
cosmological model.

In more detail: the true values of the angular diameter distance
$D_A(z)$ and Hubble expansion parameter $H(z)$ may differ from our
adopted fiducial values, $\hat{D}_A(z)$ and $\hat{H}(z)$, leading to
two scaling factors $f_\perp \equiv D_A/\hat{D}_A$ and $f_\parallel
\equiv \hat{H}/{H}$ which relate the apparent, observed tangential and
radial wavenumber components $(k_\perp',k_\parallel')$ to their true
values $(k_\perp,k_\parallel)$ via $k_\perp' = f_\perp k_\perp$ and
$k_\parallel' = f_\parallel k_\parallel$.  The apparent
two-dimensional galaxy power spectrum is then related to the
underlying fiducial galaxy power spectrum (neglecting redshift-space
distortions for the moment) by
\begin{equation}
P_{\rm g}(k',\mu') = \frac{b^2}{f_\perp^2 f_\parallel} \, P_{\rm m}
\left[ \frac{k'}{f_\perp} \sqrt{1 + \mu'^2 \left(
    \frac{f_\perp^2}{f_\parallel^2} - 1 \right)} \right]
\label{eqpkap1}
\end{equation}
(following Ballinger et al.\ 1996, Simpson \& Peacock 2010) where $k'
= \sqrt{k_\perp'^2 + k_\parallel'^2}$, $\mu' = k_\parallel'/k'$, $b$
is an unknown galaxy bias factor, and $P_{\rm m}(k)$ is the underlying
real-space, isotropic matter power spectrum.  We generated this power
spectrum shape using the CAMB software (Lewis, Challinor \& Lasenby
2000) with cosmological parameters consistent with the latest
observations of the Cosmic Microwave Background radiation (Komatsu et
al.\ 2011): matter density $\Omega_{\rm m} = 0.27$, cosmological
constant $\Omega_\Lambda = 0.73$, baryon fraction $\Omega_{\rm
  b}/\Omega_{\rm m} = 0.166$, Hubble parameter $h = 0.71$, primordial
scalar index of fluctuations $n_{\rm s} = 0.96$ and total fluctuation
amplitude $\sigma_8 = 0.8$.  We included non-linear growth of
structure in this model using the ``halofit'' prescription in CAMB
(Smith et al.\ 2003).  These assumptions do not impose a significant
model dependence in our analysis, as discussed in Section \ref{secsys}
below.

If the underlying power spectrum $P_{\rm m}(k)$ contains features at
known physical scales, which we can clearly detect as a function of
angle, then we can extract both distortion factors
$(f_\parallel,f_\perp)$.  This would be equivalent to knowing the
absolute length scale $L_0$ introduced above, and may be achieved in
principle using baryon acoustic oscillations as a standard ruler (Seo
\& Eisenstein 2003, Hu \& Haiman 2003, Glazebrook \& Blake 2005).
However, the 2D WiggleZ power spectra do not possess a sufficiently
high signal-to-noise ratio to resolve the imprint of baryon
oscillations and they do not contribute any information to the fits
presented here (although the baryon acoustic peak may be detected in
the 1D correlation function, as described by Blake et al.\ 2011b).

If the power spectrum can be well-approximated by a scale-free power
law over the range of scales of interest, which is a valid
approximation here, then we have no knowledge of $L_0$ and can only
measure $f_\parallel/f_\perp = D_A(z) H(z) / \hat{D}_A(z) \hat{H}(z) =
F/F_{\rm fid}$, where $F$ is the observed scale distortion factor and
$F_{\rm fid}$ is its value in the fiducial model.  This situation
corresponds to the Alcock-Paczynski measurement (note that any
constants in front of Equation \ref{eqpkap1} are absorbed into the
unknown bias factor $b$).  In our default implementation for
determining the best-fitting value of $F$ we varied the value of
$f_\perp = F_{\rm fid}/F$ and fixed $f_\parallel = 1$.  We checked
that our results were not significantly changed if we instead varied
$f_\parallel = F/F_{\rm fid}$, fixing $f_\perp = 1$, or varied both
$f_\perp = \sqrt{F_{\rm fid}/F}$ and $f_\parallel = \sqrt{F/F_{\rm
    fid}}$.

\subsection{Redshift-space distortion models}

The application of the Alcock-Paczynski measurement to large-scale
galaxy clustering is complicated by the fact that there is a second
physical cause of anisotropy in the observed ``redshift space'': the
coherent, bulk flows of galaxies toward clusters and superclusters
that induce systematic offsets in galaxy redshifts and hence produce a
radial (but not tangential) power spectrum distortion (Kaiser 1987,
Hamilton 1992).  In other words, the redshift-space clustering pattern
is not isotropic in the true cosmological model.

Under a set of general assumptions, independently of the cosmic
expansion history and the FRW metric, the redshift-space power
spectrum of a population of galaxies may be written as
\begin{equation}
P_{\rm g}^s(k,\mu) = P_{gg}(k) - 2 \mu^2 P_{g\theta}(k) + \mu^4
P_{\theta\theta}(k) ,
\label{eqpkz}
\end{equation}
where $P_{gg}(k) \equiv \langle |\delta_g(\vec{k})|^2 \rangle$,
$P_{g\theta}(k) \equiv \langle \delta_g(\vec{k}) \, \theta^*(\vec{k})
\rangle$ and $P_{\theta\theta}(k) \equiv \langle |\theta(\vec{k})|^2
\rangle$ are the isotropic galaxy-galaxy, galaxy-velocity and
velocity-velocity power spectra (e.g.\ Samushia et al.\ 2011).  These
equations are given in terms of $\delta_g(\vec{k})$, the Fourier
transform of the galaxy overdensity field, and $\theta(\vec{k})$, the
Fourier transform of the divergence of the peculiar velocity field in
units of the co-moving Hubble velocity (Percival \& White 2009).

We assume a linear bias factor $b$ relating the large-scale galaxy
overdensity $\delta_g$ to the underlying matter overdensity $\delta$
predicted by theory: $\delta_g = b \, \delta$.  In this case $P_{gg} =
b^2 P_{\delta\delta}$ and $P_{g\theta} = b \, P_{\delta\theta}$.
Blake et al.\ (2011a) have shown that WiggleZ galaxies follow a
distribution closely matched to the underlying matter ($b \approx 1$)
and that the galaxy-mass cross-correlation is consistent with
deterministic, linear bias over the range of scales relevant for this
analysis.

In linear perturbation theory in an FRW metric, the application of the
continuity equation implies that $P_{\delta\delta} = - f \,
P_{\delta\theta} = f^2 \, P_{\theta\theta}$, where $f$ is the growth
rate of structure, expressible in terms of the growth factor $D(a)$ at
cosmic scale factor $a$ as $f \equiv d \ln{D}/d \ln{a}$, where the
growth factor describes the evolution of the amplitude of a single
perturbation, $\delta(a) = D(a) \, \delta(1)$.  In this case we
recover the large-scale ``Kaiser limit'' (Kaiser 1987):
\begin{equation}
P_{\rm g}^s(k,\mu) = b^2 \, P_{\delta\delta}(k) \, ( 1 + \beta \,
\mu^2)^2
\label{eqkaiser}
\end{equation}
where $\beta = f/b$.  In this approximation, Equation \ref{eqpkap1} can
be modified to include redshift-space distortions as
\begin{eqnarray}
P_{\rm g}^s(k',\mu') &=& \frac{b^2}{f_\perp^2 f_\parallel} \left[ 1 +
  \mu'^2 \left( \frac{1+\beta}{F^2/F_{\rm fid}^2} - 1 \right)
  \right]^2 \nonumber \\ &\times& \left[ 1 + \mu'^2 \left(
  \frac{F_{\rm fid}^2}{F^2} -1 \right) \right]^{-2} \nonumber
\\ &\times& P_{\rm m} \left[ \frac{k'}{f_\perp} \sqrt{1 + \mu'^2
    \left( \frac{F_{\rm fid}^2}{F^2} - 1 \right)} \right]
\label{eqpkap2}
\end{eqnarray}
(following Ballinger et al.\ 1996, Simpson \& Peacock 2010).  Although
there is some degeneracy between $\beta$ and the Alcock-Paczynski
distortion parameter $F$, their angular signatures in the clustering
model are sufficiently different that both parameters may be
successfully extracted after marginalizing over the other.

However, {\it we do not use the redshift-space distortion model of
  Equation \ref{eqpkap2} in our default analysis} because the
Kaiser-limit approximation is only appropriate at the largest scales
(lowest values of $k$) due to the non-linear growth of structure at
smaller scales (Jennings et al.\ 2011, Okumura \& Jing 2011).  As a
result, fitting Equation \ref{eqpkap2} to the galaxy power spectra
could produce a systematic error in the recovered values of $F$ and
$\beta$; we include the equations simply to illustrate how $F$ and
$\beta$ may be disentangled in model-fitting.

The signature of redshift-space distortions has been modelled in more
detail by many authors using various techniques including empirical
fitting formulae (Hatton \& Cole 1998), numerical N-body dark matter
simulations (Jennings, Baugh \& Pascoli 2011) and perturbation theory
(Crocce \& Scoccimarro 2006, Taruya, Nishimichi \& Saito 2010).  A
full comparison of WiggleZ clustering data to these models was
presented by Blake et al.\ (2011a).

In the current study, we took as our fiducial redshift-space
distortion model the fitting formulae provided by Jennings et
al.\ (2011), which allow us to generate the functions
$P_{\delta\theta}(k)$ and $P_{\theta\theta}(k)$ in Equation
\ref{eqpkz} for given values of the growth rate $f$ and the non-linear
matter power spectrum $P_{\delta\delta}(k) = P_{\rm m}(k)$ (which we
obtained using ``halofit'' in CAMB).  The Jennings et al.\ (2011)
formulae have been constructed to be valid for a range of physical
dark energy models, and do not pre-suppose any specific continuity
equation linking velocity and density.  Unlike the Kaiser limit they
do not assume a Friedmann-Robertson-Walker (FRW) metric, as discussed
by Samushia et al.\ (2011).  We consider the effect of different
choices of the redshift-space distortion model in Section \ref{secsys}
below.

As a further enhancement of the redshift-space distortion model, we
also considered multiplying Equation \ref{eqpkz} by an empirical
damping function $D(k,\mu)$, representing convolution with
uncorrelated galaxy motions on small scales.  The standard choice for
this damping function is a Lorentzian $D(k,\mu) = [1 + (k_\parallel
  \sigma_v)^2]^{-1}$, where $\sigma_v$ is the pairwise velocity
dispersion, which may be varied as a free parameter but is set to zero
for our default model.  The effects of this choice on the
Alcock-Paczynski measurements are considered in Section \ref{secsys}
below.

\subsection{Model fits}

\begin{table*}
\begin{center}
\caption{This Table collects together the results of the various
  cosmological model fits to the galaxy clustering via the
  Alcock-Paczynski (AP) measurement and supernovae (SN) data presented
  in this study.  Results are presented for four different redshift
  slices.}
\label{tabres}
\begin{tabular}{cccccc}
\hline
Statistic & Method & $0.1 < z < 0.3$ & $0.3 < z < 0.5$ & $0.5 < z < 0.7$ & $0.7 < z < 0.9$ \\
\hline
Effective redshift $z$ & & $0.22$ & $0.41$ & $0.60$ & $0.78$ \\
$F_{\rm fid}(z)$ & & $0.23$ & $0.46$ & $0.71$ & $0.97$ \\
$F(z) \equiv (1+z) D_A(z) \, H(z)/c$ & AP fit & $0.28 \pm 0.04$ & $0.44 \pm 0.07$ & $0.68 \pm 0.06$ & $0.97 \pm 0.12$ \\
$f \, \sigma_8(z)$ & AP fit & $0.53 \pm 0.14$ & $0.40 \pm 0.13$ & $0.37 \pm 0.08$ & $0.49 \pm 0.12$ \\
Cross-correlation in $(F, f \, \sigma_8)$ & AP fit & $0.83$ & $0.94$ & $0.89$ & $0.84$ \\
$D_A(z) H_0/c$ & SN fit & $0.210 \pm 0.001$ & $0.376 \pm 0.005$ & $0.526 \pm 0.010$ & $0.655 \pm 0.015$ \\
$D_A(z) H_0/c$ & reconstruction & $0.209 \pm 0.001$ & $0.371 \pm 0.003$ & $0.517 \pm 0.006$ & $0.652 \pm 0.012$ \\
$H(z)/[H_0(1+z)]$ & AP + SN fit & $1.11 \pm 0.17$ & $0.83 \pm 0.13$ & $0.81 \pm 0.08$ & $0.83 \pm 0.10$ \\
$H(z)/[H_0(1+z)]$ & reconstruction & $0.91 \pm 0.02$ & $0.88 \pm 0.03$ & $0.88 \pm 0.08$ & $0.80 \pm 0.08$ \\
${\rm Om}(z)$ & reconstruction & $0.28 \pm 0.05$ & $0.29 \pm 0.07$ & $0.31 \pm 0.13$ & $0.22 \pm 0.09$ \\
$q(z)$ & reconstruction & $-0.3 \pm 0.2$ & $-0.8 \pm 0.9$ & $-1.2 \pm 2.0$ & $6.4 \pm 4.9$ \\
\hline
\end{tabular}
\end{center}
\end{table*}

\begin{figure}
\begin{center}
\resizebox{8cm}{!}{\rotatebox{270}{\includegraphics{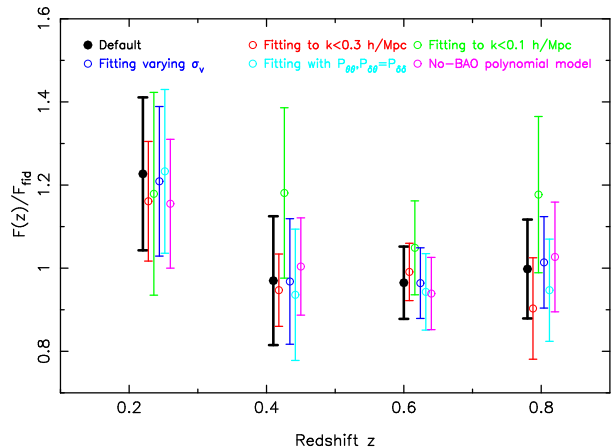}}}
\end{center}
\caption{This Figure quantifies the amplitude of systematic errors in
  our measurements of the scale distortion parameter $F(z)$ in four
  redshift slices relative to the fiducial value $F_{\rm fid}$,
  marginalized over the growth rate $f$, galaxy bias $b^2$ and
  pairwise velocity dispersion $\sigma_v$ (where appropriate).  The
  solid black data points show our default measurement using the
  redshift-space distortion model provided by Jennings et al.\ (2011)
  fitting to the wavenumber range $0 < k < k_{\rm max} = 0.2 \, h$
  Mpc$^{-1}$.  The remaining data points illustrate the effects of
  varying these assumptions: using $k_{\rm max} = 0.3 \, h$ Mpc$^{-1}$
  [red], $k_{\rm max} = 0.1 \, h$ Mpc$^{-1}$ [green], adding the
  pairwise velocity dispersion as a free parameter [blue], fitting
  using the large-scale Kaiser limit formula of Equation \ref{eqpkap2}
  [cyan], and using the data itself to define the real-space power
  spectrum via a polynomial fit to an angle-averaged measurement of
  $P(k)$ [magenta].  Each subsequent data point is slightly offset in
  redshift for clarity.}
\label{figapsys}
\end{figure}

We fitted our default Jennings et al.\ (2011) clustering model,
parameterized by $(F, f, b^2)$, to the WiggleZ Survey power spectra
over the range of scales $k < 0.2 \, h$ Mpc$^{-1}$.  Figure
\ref{figapbeta} displays the joint likelihoods of the scale distortion
parameter and growth rate in each of the redshift slices.  We quantify
the growth rate in this Figure by $f \, \sigma_8(z)$, where
$\sigma_8(z) = D(z) \, \sigma_8(0)$ quantifies the normalization of
the matter power spectrum at redshift $z$.  This is a more
model-independent observable than the growth rate itself owing to the
degeneracy between $\sigma_8(z)$ and the galaxy bias $b$ in
determining the overall galaxy clustering amplitude.  We note that
there is a strong correlation between $F$ and $f \, \sigma_8$, with
correlation coefficients $r = (0.83, 0.94, 0.89, 0.84)$ in the four
redshift slices, but both parameters may be successfully extracted.
The marginalized measurements of the growth rate are $f \, \sigma_8(z)
= (0.53 \pm 0.14, 0.40 \pm 0.13, 0.37 \pm 0.08, 0.49 \pm 0.12)$.
These measurements are consistent with those reported by Blake et
al.\ (2011a) who considered fits of redshift-space distortion models
assuming $F = F_{\rm fid}$.  The resulting measurements of $F$,
marginalizing over $f$ and $b^2$, are plotted in Figure \ref{figapsys}
as the solid black data points.  Our result is $F(z) \equiv (1+z)
D_A(z) \, H(z)/c = (0.28 \pm 0.04, 0.44 \pm 0.07, 0.68 \pm 0.06, 0.97
\pm 0.12)$ in the four redshift slices.  All of the different
measurements reported in Sections \ref{secap}, \ref{secapsn} and
\ref{secrecon} are collected together in Table \ref{tabres} for
convenience.

\subsection{Systematics tests}
\label{secsys}

Figure \ref{figapsys} presents some tests of the sensitivity of our
results to the different assumptions in the model described above.
Firstly, we checked that our measurement of $F$ was not significantly
changed if we replaced our default model with a range of other
implementations of redshift-space distortions discussed in the recent
literature, including generating the density and velocity power
spectra using Renormalized Perturbation Theory (Crocce \& Scoccimarro
2006) and adding the correction terms proposed by Taruya et
al.\ (2010).  Indeed, using the simplication of Equation
\ref{eqpkap2}, where $P_{\delta\delta}(k)$ is the non-linear power
spectrum obtained from CAMB, produced an unchanged result for $F$
(within the statistical errors), as illustrated in Figure
\ref{figapsys} (the cyan data points).

We also considered including and excluding the pairwise velocity
dispersion as a free parameter.  If we fit for $F$ by varying
$f_\perp$ and fixing $f_\parallel$, then marginalizing over the
additional free parameter $\sigma_v$ makes no difference to our
results (illustrated by the blue data points in Figure
  \ref{figapsys}).  However, if we fit for $F$ by varying
$f_\parallel$ and fixing $f_\perp$ then the resulting error in $F$ is
significantly increased by marginalizing over $\sigma_v$, due to the
cross-talk between $k_\parallel$ and $\sigma_v$ in the damping term.

Any potential systematic error in the redshift-space distortion model
is likely to worsen at smaller scales (higher values of $k$) where the
modelling becomes less accurate.  We carefully compared different
choices of the fitting range $0 < k < k_{\rm max}$, in order to ensure
that any systematic error in the derived distortion parameter was not
significant.  The results of some of these tests are plotted in Figure
\ref{figapsys}.  Our fiducial choice, $k_{\rm max} = 0.2 \, h$
Mpc$^{-1}$, was determined by the consideration that the measurements
of $F$ should not differ systematically between different
implementations of the redshift-space distortion model.  Results for
$k_{\rm max} = 0.1$ and $0.3 \, h$ Mpc$^{-1}$ are displayed as the red
and green data points in Figure \ref{figapsys}.  As a further test we
repeated the measurements of $F$ for independent scale ranges
$k=0-0.1$, $0.1-0.2$ and $0.2-0.3 \, h$ Mpc$^{-1}$; the results were
consistent within the statistical errors.

Our analysis uses an underlying isotropic matter power spectrum that
is consistent with the latest observations of the Cosmic Microwave
Background radiation (Komatsu et al.\ 2011).  In order to show that
this does not introduce any sensitivity to the model into our analysis
we checked that our results were unchanged if we instead generated a
function proportional to $P_{\rm m}(k)$ using a polynomial fit to the
observed, angle-averaged galaxy power spectrum in the fiducial model;
this comparison is shown as the magenta data points in Figure
\ref{figapsys}.  The similarity between these measurements and our
fiducial results provides further evidence that the baryon acoustic
oscillations (which do not appear in a smooth polynomial model) are
not contributing any information to the Alcock-Paczynski distortion
fits.

We conclude from these tests that the systematic error in $F$ induced
from modelling redshift-space distortions is much lower than the
statistical error in the measurement.  Our results are therefore
insensitive to the model adopted (to first order), even though a
series of model assumptions are necessary to produce these fits.

\section{Determination of the cosmic expansion history}
\label{secapsn}

We converted our Alcock-Paczynski measurements of the scale distortion
parameter $F(z) = (1+z) D_A(z) H(z)/c$ into a determination of the
cosmic expansion history $H(z)$ by using Type Ia supernovae (SNe Ia)
data to fix the cosmic distance-redshift relation.  We used the
``Union-2'' compilation by Amanullah et al.\ (2010) as our supernovae
dataset, obtained from the website {\tt
  http://supernova.lbl.gov/Union}.  This compilation of 557 supernovae
includes data from Hamuy et al.\ (1996), Riess et al.\ (1999, 2007),
Astier et al.\ (2006), Jha et al.\ (2006), Wood-Vasey et al.\ (2007),
Holtzman et al.\ (2008), Hicken et al.\ (2009) and Kessler et
al.\ (2009).

\begin{figure}
\begin{center}
\resizebox{8cm}{!}{\rotatebox{270}{\includegraphics{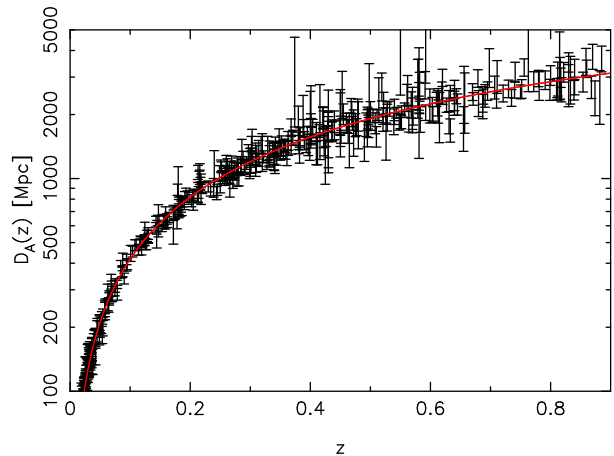}}}
\end{center}
\caption{This Figure displays the best-fitting 3rd-order polynomial to
  the Union-2 compilation of supernovae data, normalized as a plot of
  angular-diameter distance versus redshift.}
\label{figdasn}
\end{figure}

Given that the normalization of the supernova Hubble diagram $M - 5 \,
{\rm log}_{10} h$ is treated as an unknown parameter, the supernovae
data yields the relative luminosity distance $D_L(z) \, H_0/c$.  We
determined a model-independent value of this quantity in each redshift
slice by fitting the distance-redshift relation as a 3rd-order
polynomial in $z$ (over the redshift range $0 < z < 0.9$) and
marginalizing over the values of the polynomial coefficients.  We used
the full covariance matrix of the supernovae measurements including
systematic errors, and checked that our results were not significantly
changed by assuming a 2nd-order or 4th-order polynomial instead.  Our
results at the four redshifts $z = (0.22, 0.41, 0.6, 0.78)$ were
$D_A(z) \, H_0/c = (0.210 \pm 0.001, 0.376 \pm 0.005, 0.526 \pm 0.010,
0.655 \pm 0.015)$, where we have converted the luminosity distances to
angular diameter distances assuming $D_L(z) = D_A(z) (1+z)^2$.  We
note that the inclusion of the supernovae systematics covariance
matrix (compared to uncorrelated errors excluding systematics)
increases the errors in these measurements of $D_A \, H_0/c$ by a
factor of two.  The supernovae data points and best-fitting 3rd-order
polynomial model are displayed in Figure \ref{figdasn}.  This model
provides a good fit to the data, with a chi-squared statistic of
$486.6$ for $519$ degrees of freedom.

\begin{figure}
\begin{center}
\resizebox{8cm}{!}{\rotatebox{270}{\includegraphics{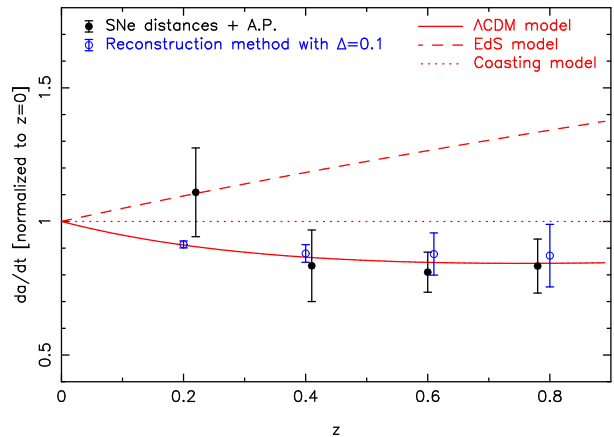}}}
\end{center}
\caption{This Figure displays our measurement of the evolution of the
  cosmic expansion rate using Alcock-Paczynski and supernovae data.
  The expansion rate is displayed using the value of
  $\dot{a}/\dot{a}_0 = H(z)/[H_0(1+z)]$; accelerating expansion
  implies a decrease in the value of this quantity with increasing
  redshift.  The black data points are obtained by combining
  Alcock-Paczynski measurements of $(1+z) D_A(z) H(z)/c$ in four
  independent redshift slices with supernovae distance determinations
  of $D_L(z) H_0/c$ at these redshifts, and are independent of
  curvature.  The thicker, blue data points result from applying the
  distance reconstruction method of Shafieloo et al.\ (2006) to both
  the supernovae and Alcock-Paczynski data, producing optimal errors
  at both low and high redshift but making the additional assumptions
  of zero spatial curvature and that $D_A(z)$ may be expressed in
  terms of an integral over $1/H(z)$.  Predictions are plotted for
  three different models: a fiducial $\Lambda$CDM model with
  $\Omega_{\rm m} = 0.27$ (solid line), an Einstein de-Sitter model
  with $\Omega_{\rm m} = 1$ (dashed line), and a ``coasting'' model
  where $\dot{a} =$ constant (dotted line).}
\label{figadotmeas}
\end{figure}

Combining the Alcock-Paczynski and supernovae measurements in the four
redshift slices we find that $H(z)/[H_0(1+z)] \equiv \dot{a}/\dot{a}_0
= (1.11 \pm 0.17, 0.83 \pm 0.13, 0.81 \pm 0.08, 0.83 \pm 0.10)$ where
$\dot{a}_0 \equiv H_0$.  These results are plotted as the black error
bars in Figure \ref{figadotmeas} and compared to three expansion
history models: a $\Lambda$CDM cosmology with matter density
$\Omega_{\rm m} = 0.27$, an Einstein-de-Sitter (EdS) model with
$\Omega_{\rm m} = 1$ and a ``coasting'' model for which $da/dt = H_0$
at all times.  We find that our measurements are consistent with the
$\Lambda$CDM model.  Figure \ref{figadotmeas} constitutes our
model-independent measurement of the cosmic expansion rate $da/dt$,
with $10$-$15\%$ precision in each of the four redshift slices.  We
note that the systematic errors in the supernovae measurements are
insignificant in the determination of $H(z)/[H_0(1+z)]$ because the
error propagation is dominated by the error in the Alcock-Paczynski
measurement.

Accelerating expansion implies a decrease in the value of $da/dt$ with
increasing redshift.  This is directly observed in our data: if we fit
for a constant value of $\dot{a}/\dot{a}_0$ we find that the range
$\dot{a}/\dot{a}_0 < 1$, containing accelerating-expansion models,
contains $99.86\%$ of the probability.  Hence in this assessment,
accelerating-expansion models are preferred with a statistical
significance of more than 3-$\sigma$.  We stress that this is a
non-parametric result, insensitive to the redshift-space distortion
model adopted, based solely on combining Alcock-Paczynski measurements
of $D_A(z) H(z)$ with supernovae measurements of $D_A(z)$

\section{Full distance-redshift reconstruction}
\label{secrecon}

\begin{figure*}
\begin{center}
\resizebox{16cm}{!}{\rotatebox{270}{\includegraphics{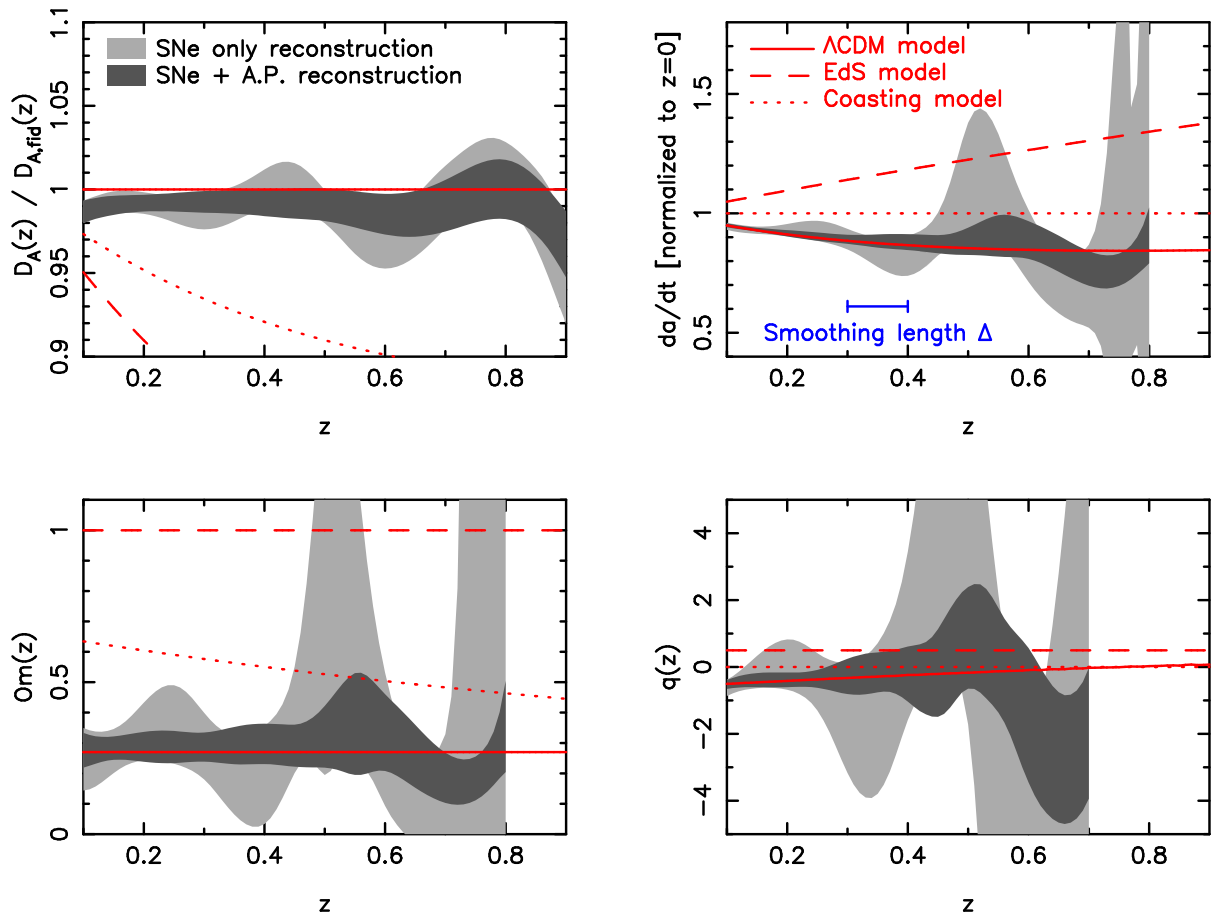}}}
\end{center}
\caption{This Figure shows our non-parametric reconstruction of the
  cosmic expansion history using Alcock-Paczynski and supernovae data.
  The four panels of this figure display our reconstructions of the
  distance-redshift relation $D_A(z)$, the expansion rate
  $\dot{a}/H_0$, the ${\rm Om}(z)$ statistic and the deceleration
  parameter $q(z)$ using our adaptation of the iterative method of
  Shafieloo et al.\ (2006) and Shafieloo \& Clarkson (2010).  The
  distance-redshift relation in the upper left-hand panel is divided
  by a fiducial model for clarity, where the model corresponds to a
  flat $\Lambda$CDM cosmology with $\Omega_{\rm m} = 0.27$.  This
  fiducial model is shown as the solid line in all panels; Einstein
  de-Sitter and coasting models are also shown defined as in Figure
  \ref{figadotmeas}.  The shaded regions illustrate the $68\%$
  confidence range of the reconstructions of each quantity obtained
  using bootstrap resamples of the data.  The dark-grey regions
  utilize a combination of the Alcock-Paczynski and supernovae data
  and the light-grey regions are based on the supernovae data alone.
  The redshift smoothing scale $\Delta = 0.1$ is also illustrated.
  The reconstructions in each case are terminated when the SNe-only
  results become very noisy; this maximum redshift reduces with each
  subsequent derivative of the distance-redshift relation [i.e.\ is
    lowest for $q(z)$].}
\label{figadotrecon}
\end{figure*}

In this Section we use our datasets to map out the full cosmic
expansion history via the non-parametric reconstruction method
introduced by Shafieloo et al.\ (2006).  We note that, unlike the
preceding analysis, this approach assumes zero spatial curvature and
that $D_A(z)$ can be expressed in terms of an integral over $1/H(z)$
(i.e., the FRW metric).  Hence we expect to derive tighter constraints
based on more assumptions.

\subsection{Method for reconstructing $D_A(z)$}

We first describe the application of the Shafieloo et al.\ (2006)
distance reconstruction method to SNe Ia data.  We then outline its
generalization to enable the inclusion of the Alcock-Paczynski
measurements.

The methodology of Shafieloo et al.\ (2006) is an iterative approach
that deduces a distance-redshift curve $D_A(z)$ from an initial guess
model by smoothing the residuals between the model and measurements of
$D_A(z)$ and using these residuals to modify the guess.  Iteration
proceeds until the chi-squared statistic between model and data is
minimized.  The only parameter required by the method is the width of
the smoothing kernel in redshift, $\Delta$ (our fiducial value is
$\Delta = 0.1$, other choices are considered below).  For the initial
guess model to commence the iteration, we chose a flat $\Lambda$CDM
model with $\Omega_{\rm m} = 0.27$.  Our results have no sensitivity
to the choice of initial guess (as illustrated by the offset between
the reconstructed and fiducial distance curve in the top left-hand
panel of Figure \ref{figadotrecon}).  We defined the model on a grid
of redshift steps spaced by $dz=0.01$ from $z=0$ to $z=1$.

The equation generating iteration $i+1$ from iteration $i$, in terms
of the supernovae data points $j$, is given in Shafieloo \& Clarkson
(2010):
\begin{eqnarray}
& & \ln{D_L(z)}^{i+1} = \ln{D_L(z)}^i + N(z) \times \nonumber \\ & &
  \sum_j \frac{[\ln{D_L(z_j)} - \ln{D_L(z_j)}^i]}{\sigma_j^2} \exp{
    \left[ - \frac{(z_j - z)^2}{2\Delta^2} \right] }
\label{eqsniter}
\end{eqnarray}
where the redshift-dependent normalization is
\begin{equation}
N(z)^{-1} = \sum_j \frac{1}{\sigma_j^2} \exp{ \left[ -
    \frac{(z_j - z)^2}{2\Delta^2} \right] }
\end{equation}
For the inverse-variance weighting $\sigma_j$ we followed Shafieloo \&
Clarkson (2010) and used the error in each supernova luminosity
distance.

We then modified the distance reconstruction method to utilize the
residuals with respect to the Alcock-Paczynski data points.  We
initially generated an iterative correction to the quantity $F(z) =
(1+z) D_A(z) H(z)/c$ using an analogous method to Equation
\ref{eqsniter}:
\begin{equation}
\Delta F(z) = M(z) \sum_k \frac{F(z_k) - F(z_k)^i}{\sigma_k^2} \exp{
  \left[ - \frac{(z_k-z)^2}{2\Delta^2} \right] }
\end{equation}
where $k$ labels the Alcock-Paczynski measurements at redshifts $z_k$,
with errors $\sigma_k$ in $F(z)$, and the normalization factor is
given by
\begin{equation}
M(z)^{-1} = \sum_k \frac{1}{\sigma_k^2} \exp{ \left[ -
    \frac{(z_k-z)^2}{2\Delta^2} \right] }
\end{equation}
We then converted the iteration in $F(z)$ into an iteration in $D_L(z)
= (1+z)^2 D_A(z)$ using
\begin{eqnarray}
D_A(z)^{i+1} &=& D_A(z)^i \nonumber \\ &-& \sum_{z'} (1+z') D_A(z')^i
\frac{\Delta F(z')}{[F(z')^i]^2} \Delta z' ,
\label{eqapiter}
\end{eqnarray}
where $\Delta z' = 0.01$ is the redshift interval of the gridding of
the model curves.  The right-hand side of Equation \ref{eqapiter} is
derived by expressing $D_A(z) = \sum_{z'} [c/H(z')] \Delta z'$,
substituting $c/H(z') = (1+z') D_A(z')/F(z')$, and taking the
derivative with respect to $F$.  At each step in the iteration we
applied the residual distance correction for both the supernovae and
Alcock-Paczynski data points, proceeding until the chi-squared
statistic was minimized.

We estimated the error in the distance-redshift reconstruction by
repeating the iterative process using bootstrap re-samples of the
datasets.  The range of resulting distance-redshift curves across the
re-samples defines the error in the reconstruction.  We performed
extensive tests using Monte Carlo realizations of synthetic datasets
to verify that this procedure produced bias-free reconstructions with
reliable error ranges.  We note that this procedure differs from that
of Shafieloo et al.\ (2006) and Shafieloo \& Clarkson (2010), who used
the set of distance curves $[\ln{D_L(z)}^i]$ generated by a single
application of the iteration process of Equation \ref{eqsniter} to
estimate the error in the reconstruction.  In order to obtain the
error in the reconstructed curves we constructed separate bootstrap
re-samples of the supernovae and Alcock-Paczynski data points.

\subsection{Method for reconstructing $H(z)$, ${\rm Om}(z)$ and $q(z)$}

We used the reconstructed distance-redshift curves to determine three
other diagnostics that describe the cosmic expansion history.
Firstly, assuming a flat Universe we can differentiate the
reconstructed distance-redshift curves to determine the expansion rate
$\dot{a}/\dot{a}_0 = H(z)/[H_0(1+z)]$.  Secondly, a useful diagnostic
of the expansion history that can be derived from $H(z)$ is the ``Om''
statistic (Sahni, Shafieloo \& Starobinsky 2008):
\begin{equation}
{\rm Om}(z) \equiv \frac{[H(z)/H_0]^2 - 1}{(1+z)^3 - 1} .
\end{equation}
In a spatially-flat $\Lambda$CDM model this statistic is constant at
different redshifts and equal to today's value of the matter density
parameter $\Omega_{\rm m}$.  In universes with different curvature, or
containing dark energy with different properties to a cosmological
constant, ${\rm Om}(z)$ would evolve with redshift.  Finally, by a
second differentiation of the distance-redshift curves we can obtain
the deceleration parameter $q(z) \equiv -\ddot{a} a/\dot{a}^2$:
\begin{equation}
q(z) = (1+z) \frac{dH(z)/dz}{H(z)} - 1 .
\end{equation}
The confidence ranges for the reconstruction of each quantity are
established using bootstrap re-sampling.

\subsection{Results of the reconstruction}

Figure \ref{figadotrecon} displays our non-parametric reconstructions
of the four quantities $D_A(z)$, $da/dt$, ${\rm Om}(z)$ and $q(z)$.
The light-grey regions display the $68\%$ confidence range of
reconstructions using the supernovae data alone, and the inner
dark-grey region derives from the combination of supernovae and
Alcock-Paczynski measurements.

The addition of the Alcock-Paczynski data does not strongly enhance
the reconstruction of the distance-redshift relation $D_A(z)$, which
is very well-measured by supernovae alone.  However, the direct
sensitivity of $F(z)$ to $H(z)$ produces dramatic improvements in the
reconstruction of the expansion-rate history, with a factor of 3
shrinking of the confidence ranges in $H(z)$ for $z > 0.2$.  At the
four redshifts $z = (0.22, 0.41, 0.6, 0.78)$ the reconstruction method
gives us $H(z)/[H_0(1+z)] = (0.91 \pm 0.02, 0.88 \pm 0.03, 0.88
\pm 0.08, 0.80 \pm 0.08)$.  These measurements are correlated, but
the level of correlation is low because the fiducial choice of the
smoothing length $\Delta = 0.1$ is somewhat smaller than the bin width
$\Delta z = 0.2$.

Our measurements establish the constancy of ${\rm Om}(z)$ with
redshift for the first time with reasonable precision, a key
prediction of a flat $\Lambda$CDM model.  At the four redshifts listed
above the reconstruction method produces ${\rm Om}(z) = (0.28 \pm
0.05, 0.29 \pm 0.07, 0.31 \pm 0.13, 0.22 \pm 0.09)$.  If we fit these
values for the matter density in a flat $\Lambda$CDM model we find
$\Omega_{\rm m} = 0.29 \pm 0.03$, consistent with fits to the CMB
radiation (Komatsu et al.\ 2011).

We also overplot in Figure \ref{figadotmeas} the non-parametric
measurements of $da/dt$ obtained by this reconstruction method in the
four redshift slices.  Comparison of the errors with the direct
combination of supernovae distances and Alcock-Paczynski data shows
that at low redshift the supernovae data alone place the most powerful
constraint on the expansion history, greatly improving the results
presented in Section \ref{secapsn}, but at higher redshifts the
Alcock-Paczynski data are crucial in order to obtain accurate
measurements.

\begin{figure*}
\begin{center}
\resizebox{16cm}{!}{\rotatebox{270}{\includegraphics{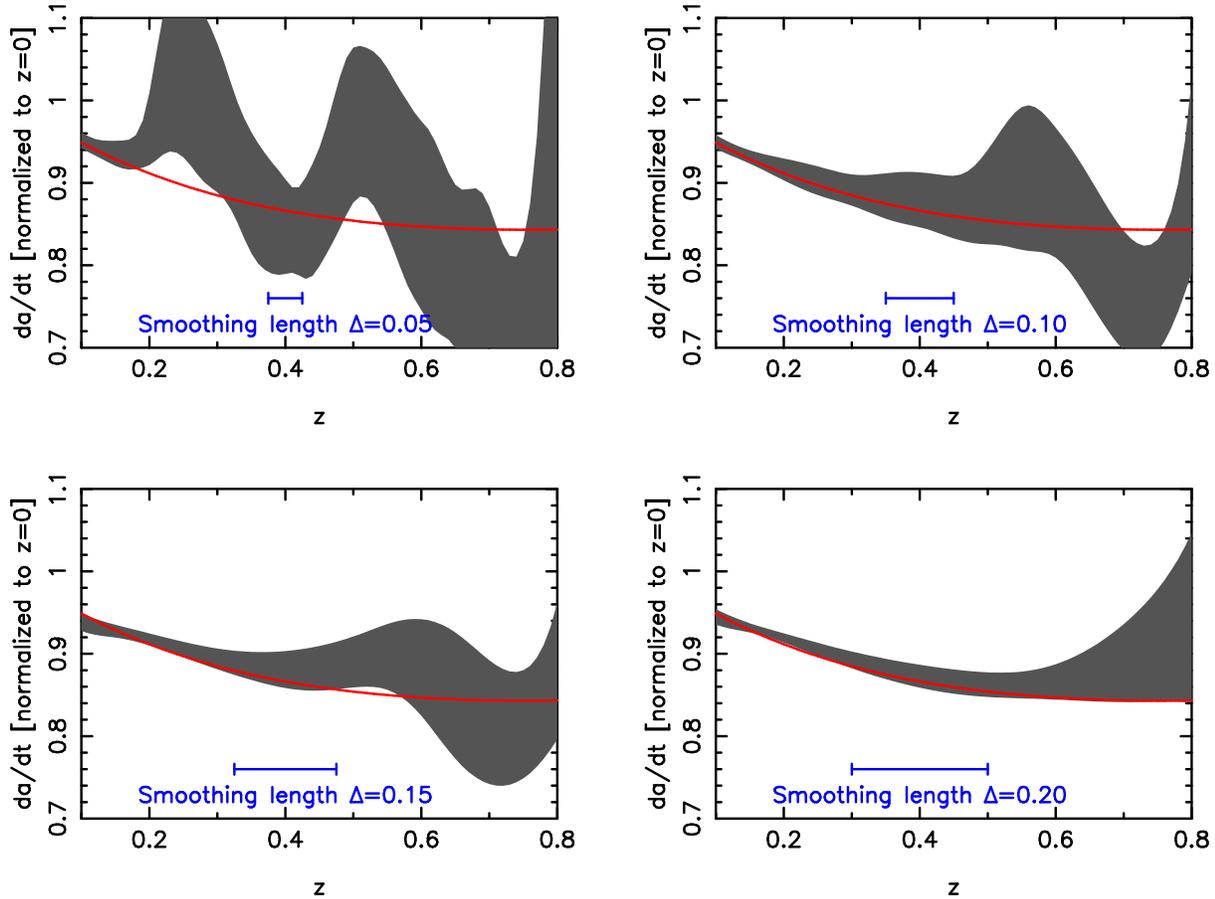}}}
\end{center}
\caption{In this Figure we repeat the reconstruction of
$\dot{a}/H_0$ using both supernovae and Alcock-Paczynski data for
different choices of the smoothing length $\Delta$, the only free
parameter in the reconstruction algorithm.  The solid band shows the
$68\%$ confidence range spanned by the reconstructions and the solid
red line is the prediction of a flat $\Lambda$CDM model with
$\Omega_{\rm m} = 0.27$.}
\label{figreconsmooth}
\end{figure*}

Figure \ref{figreconsmooth} illustrates the effect of changing the
smoothing length on the reconstructed expansion history
$\dot{a}/\dot{a}_0$, considering values $\Delta = 0.05, 0.10, 0.15$
and $0.20$.  A flat $\Lambda$CDM model with $\Omega_{\rm m} = 0.27$,
shown as the solid line, continues to be a good description of the
expansion history regardless of the smoothing scale.  We checked that
the error in fitting $\Omega_{\rm m}$ to these data was insensitive to
$\Delta$ when covariances between different redshifts were taken into
account (estimating these covariances using the bootstrap resamples).

\section{Conclusions}
\label{secconc}

We summarize our study as follows:

\begin{itemize}

\item We have performed joint fits for the Alcock-Paczynski scale
  distortion parameter $F(z) = (1+z) D_A(z) H(z)/c$ and the normalized
  growth rate $f \, \sigma_8(z)$, using 2D power spectra measured from
  the WiggleZ Dark Energy Survey in four redshift bins in the range
  $0.1 < z < 0.9$.  We can separate the contributions of
  Alcock-Paczynski and redshift-space distortions, and our result is
  insensitive to the range of models we examined for non-linear
  redshift-space effects when fitted to the wavenumber range $k < 0.2
  \, h$ Mpc$^{-1}$.

\item By combining the Alcock-Paczynski fits with luminosity-distance
  measurements using Type Ia supernovae, we determined the cosmic
  expansion rate $\dot{a}/\dot{a}_0 = H(z)/[H_0(1+z)]$ in four
  redshift slices with $10$-$15\%$ accuracy.  Our results for
  redshifts $z = (0.22, 0.41, 0.6, 0.78)$ are $\dot{a}/\dot{a}_0 =
  (1.11 \pm 0.17, 0.83 \pm 0.13, 0.81 \pm 0.08, 0.83 \pm 0.10)$.
  These measurements are independent of spatial curvature.

\item Our measurements show that the value of $\dot{a}$ was lower at
  high redshift, demonstrating that the cosmic expansion has
  accelerated.  If we fit for a constant value of $\dot{a}$ in the
  redshift range $0.1 < z < 0.9$ we find that accelerating-expansion
  models are favoured with a statistical significance of more than
  3-$\sigma$.

\item We used an adapted version of the reconstruction method of
  Shafieloo et al.\ (2006) to model the continuous cosmic expansion
  history since $z=0.9$ using both the Alcock-Paczynski and supernovae
  datasets.  The Alcock-Paczynski measurements enable a much more
  accurate determination of the expansion history than the supernovae
  data alone.  We demonstrate that the quantity ${\rm Om}(z) \equiv
  \{[H(z)/H_0]^2-1\}/[(1+z)^3-1]$ is constant with redshift, as
  expected in a spatially-flat $\Lambda$CDM model; fitting for the
  value of this quantity allows us produce an estimate of the
  matter-density parameter $\Omega_{\rm m} = 0.29 \pm 0.03$,
  consistent with fits to the CMB radiation (Komatsu et al.\ 2011).

\end{itemize}

We conclude that accelerating cosmic expansion can be recovered from
cosmological data in a non-parametric manner and hence is a real
physical phenomenon that must be accounted for by theory.  Our results
are independent of the cosmological model adopted, although our data
analysis assumes a homogeneous Universe.  The measured expansion
history is well-fit by a cosmological constant which grows in relative
importance with cosmic time.  The combination of Alcock-Paczynski and
supernovae data is a novel approach that enables direct observation of
the Hubble expansion rate and will be strengthened in the future by
the availability of new galaxy redshift survey data.

\section*{Acknowledgments}

We thank the anonymous referee for helpful comments.  CB acknowledges
useful discussions with Berian James, Juliana Kwan and Arman
Shafieloo.  We acknowledge financial support from the Australian
Research Council through Discovery Project grants DP0772084 and
DP1093738 and Linkage International travel grant LX0881951.  SC
acknowledges the support of an Australian Research Council QEII
Fellowship.  MJD and TMD thank the Gregg Thompson Dark Energy Travel
Fund for financial support.  GALEX (the Galaxy Evolution Explorer) is
a NASA Small Explorer, launched in April 2003.  We gratefully
acknowledge NASA's support for construction, operation and science
analysis for the GALEX mission, developed in co-operation with the
Centre National d'Etudes Spatiales of France and the Korean Ministry
of Science and Technology.  We thank the Anglo-Australian Telescope
Allocation Committee for supporting the WiggleZ survey over 9
semesters, and we are very grateful for the dedicated work of the
staff of the Australian Astronomical Observatory in the development
and support of the AAOmega spectrograph, and the running of the AAT.

\end{document}